\documentclass{sig-alternate-05-2015}

\usepackage[linesnumbered]{algorithm2e}
\usepackage[caption=false]{subfig}
\usepackage{graphicx}
\usepackage{amsfonts}
 \usepackage{multirow}
 \usepackage{slashbox}

\numberofauthors{1} \author{
\alignauthor
Brian Brost\textsuperscript{1}, Ingemar J. Cox\textsuperscript{1,2}, Yevgeny Seldin\textsuperscript{1}, Christina Lioma\textsuperscript{1} \\
       \affaddr{ \textsuperscript{1} University of Copenhagen}\\
       \affaddr{ \textsuperscript{2} University College London}\\
       \email{brian.brost@di.ku.dk, ingemar@ieee.org, seldin@di.ku.dk, c.lioma@di.ku.dk}
}

\title{An Improved Multileaving Algorithm for Online Ranker Evaluation}

\date{30 July 1999}

\CopyrightYear{2016} 
\setcopyright{acmlicensed}
\conferenceinfo{SIGIR '16,}{July 17 - 21, 2016, Pisa, Italy}
\isbn{978-1-4503-4069-4/16/07}\acmPrice{\$15.00}
\doi{http://dx.doi.org/10.1145/2911451.2914706}

\begin{document}

\maketitle

\begin{abstract}
Online ranker evaluation is a key challenge in information retrieval. An important task in the online evaluation of rankers is using implicit user feedback for inferring preferences between rankers. Interleaving methods have been found to be efficient and sensitive, i.e. they can quickly detect even small differences in quality. It has recently been shown that multileaving methods exhibit similar sensitivity but can be more efficient than interleaving methods.
This paper presents empirical results demonstrating that existing multileaving methods either do not scale well with the number of rankers, or, more problematically, can produce results which substantially differ from evaluation measures like NDCG.
The latter problem is caused by the fact that they do not correctly account for the similarities that can occur between rankers being multileaved. We propose a new multileaving method for handling this problem and demonstrate that it substantially outperforms existing methods, in some cases reducing errors by as much as 50\%. 
\end{abstract}

\section{Introduction}
\label{sec:ProblemSetting}

Online evaluation using interleaving is an increasingly popular paradigm in ranker evaluation and has been found to be efficient and sensitive \cite{chapelle2012large}. Here efficient means that relatively little click feedback is required to reliably distinguish rankers, and sensitive means that interleaving can distinguish between rankers with very similar retrieval quality. In addition to these requirements an important criterion for evaluating interleaving methods is that they be unbiased, in the sense that they do not systematically favour certain rankers independently of their actual quality.

Multileaving methods were recently introduced \cite{schuth2015probabilistic, schuth2014multileaved} and potentially offer substantial improvements over interleaving in terms of efficiency, since they can compare sets of rankers of arbitrary size at each comparison. They were also found to be similarly sensitive to interleaving \cite{schuth2014multileaved}.

Section~\ref{sec:related} describes related work, including two state of the art methods, Team Draft Multileave (TDM) and Probabilistic Multileave (PM). We observe that TDM does not scale well as the number of rankers in the comparison set increases, limiting its efficiency. We show that PM can fail to properly account for ranker similarities, introducing bias. Section~\ref{sec:SOSM} then describes our proposed solution and Section~\ref{sec:experiment} provides experimental results demonstrating the new algorithm's superiority. Our contributions are to produce a new multileaving method which outperforms current methods and to identify that PM can be biased.


\section{Related Work}
\label{sec:related}

Multileaving consists of two stages. First we create the multileaved list by sampling from the individual ranked lists, and then we credit rankers based on user clicks on the multileaved list and thereby infer a preference ordering of the rankers. Three multileaving methods have been proposed, namely, Team Draft Multileave (TDM) \cite{schuth2014multileaved}, Optimised Multileave \cite{schuth2014multileaved} and Probabilistic Multileave (PM) \cite{schuth2015probabilistic}. We describe the two best performing methods, TDM and PM.

TDM creates the multileaved list in rounds. In each round a random ordering of the rankers is decided and the top document that has not yet appeared in the multileaving is drawn from each ranker. This process is then repeated until the multileaved list is of sufficient length. In the credit inference stage of TDM, rankers are credited for each click on a document drawn from the corresponding ranker. This credit does not consider the position of the document in the ranker's retrieved list. A matrix, $M$, of pair-wise preferences between pairs of rankers is then inferred based on which ranker was given more credit.

Since rankers are only credited for clicks on documents drawn from the corresponding ranker, TDM does not scale well to comparing more rankers than there are documents in the multileaved list. Since users of search engines typically only inspect the first results page, the multileaved list is effectively only of length 10. We are often interested in comparing significantly more than just 10 rankers, so there is a need for multileaving methods which scale better to larger comparison sets. 

PM also creates the multileaved list in rounds. In each round a random ordering of the rankers is decided. Then, a document is probabilistically selected from the ranker, where the probability of drawing a document $d$ from ranker $R_j$ is determined solely by the document's rank and is given by Equation~\ref{eq1}, where $r_j(d)$ is the rank of document $d$ in ranker $R_j$, and $D$ is the set of documents ranked by $R_j$. 
\begin{equation}
P(d | R_j)=\frac{\frac{1}{r_j(d)^3}}{\sum_{d' \in D}\frac{1}{r_j(d')^3}}
\label{eq1}
\end{equation}
Note that when a document is drawn, the document is removed from all the rankers' retrieved lists. In the next round, the probabilities of the remaining documents are recalculated according to Equation~\ref{eq1}, where the rankings, $r_j(d)$, are now determined in the absence of previously chosen documents. 

In the credit inference stage, PM considers all possible assignments of documents to rankers that could have occurred, and weight each assignment based on its probability. An assignment, $a$, has probability, $P(a)$, given by
\begin{equation}
P(a) = \prod_{r=1}^{L}P(d_r|R_{\alpha(r)})P(R_{\alpha(r)})
\end{equation}
where $L$, is the length of the multileaved list, $P(d_r|R_{\alpha(r)})$, is the probability of drawing document $d_r$ from the assigned ranker, $R_{\alpha(r)}$ and is given by Equation~\ref{eq1}, and $P(R_{\alpha(r)})$ is given by $1/|R|$. For an assignment, $a$, Ranker $R_j$ is given credit, $o_j(a)$ equal to the number of assigned documents clicked on. The total credit, $o_j(A)$, assigned to ranker $R_j$, is given by 
$o_j(A) = \sum_{a \in A} o_j(a)P(a)$, where $A$ is the set of all possible assignments. A matrix, $M$, of pair-wise preferences between pairs of rankers is then inferred based on which ranker was given more credit.

PM can be biased since rankers can benefit from the presence of documents contributed by similar rankers, and these rankers will therefore perform better according to PM than they actually do in practice. To illustrate this problem, consider the following simple example: three rankers and their corresponding retrieved lists:
$(R_1: D_1, D_2)$
$(R_2: D_2, D_1)$, and
$(R_3: D_2, D_1)$.
The possible multileavings of length two, are $\{D_1,D_2\}$ and $\{D_2,D_1\}$. The former multileaving occurs with probability 0.3704, and the latter occurs with probability 0.6296. Assume that $D_1$ and $D_2$ are both relevant and always clicked on, i.e. all three rankers have equal performance. Even though the rankers are equally good, $R_1$ will lose to the other two rankers with probability $0.6296$ due to the fact that when the multileaving $\{D_1,D_2\}$ occurs, $R_1$ is given more credit in the credit inference stage of PM and if $\{D_2,D_1\}$ occurs, $R_2$ and $R_3$ are given more credit. Thus, for PM, the presence of similar rankers introduces bias and distorts the outcome of comparisons. Similar rankers benefit because their assignments are weighted higher. 

\section{Sample-only Scored Multileave}
\label{sec:SOSM}

We propose a multileaving method called Sample-only Scored Multileave (SOSM) which scales well with the number of rankers being compared, without introducing bias. The main difference relative to PM is that the score attributed to each ranker {\em only} depends on how each ranker ranks the sample of documents contained in the multileaving, {\em not} how each ranker ranks documents in their original retrieved lists. In this way, if a document is preferentially sampled, it will not disproportionally disadvantage other rankers provided they rank the sample well. 

The process of creating the multileaved list in SOSM is identical to that in TDM described in Section~\ref{sec:related}.

To infer preferences, each ranker ranks the documents in the multileaved list such that $r^\prime_j(d)$ denotes the order of document $d$ in the multileaved list according to ranker $R_j$. Letting $D_M$ denote the documents of the multileaved list, the score of document $d$ for ranker $R_j$ is given in Equation~\ref{eq2}. Letting $C$ denote the clicked documents, ranker $R_j$ is credited with $\sum_{d \in C} s(d | R_j)$, where
\begin{equation}
s(d | R_j)=\frac{\frac{1}{r^\prime_j(d)^3}}{\sum_{d' \in D_M}\frac{1}{r^\prime_j(d')^3}}
\label{eq2}
\end{equation}
A matrix, $M$, of preferences between pairs of rankers is then inferred based on which ranker was given more credit. Note the similarity between our scoring function in Equation~\ref{eq2}, and that of Equation~\ref{eq1} used by PM. The only difference is that in the denominator we only sum over the documents contained in the multileaved list. 

SOSM scales well with the number of rankers being compared, as verified experimentally in Section~\ref{sec:experiment}, and is also unbiased. It is simple to verify that SOSM is unbiased, according to the definition of bias given in \cite{hofmann2013fidelity}. Additionally, we verify the unbiasedness of SOSM experimentally in Section~\ref{sec:experiment}.

\section{Experimental Evaluation}
\label{sec:experiment}

In our problem setup, we are given a set of rankers $R$ whose performance we want to evaluate on a dataset using click feedback \cite{schuth2014multileaved}. Multileaving methods output an $|R| \times |R|$ matrix $M$ after each comparison, where $M_{ij}$ is 1 if the multileaving method inferred a preference for ranker $R_i$ over $R_j$, 0 if it inferred a preference for ranker $R_j$ over $R_i$ and $0.5$ if no preference was inferred between the rankers. We then define $\hat{M}(t)$ such that $\hat{M}_{ij}$ is the average over $t$ multileaved comparisons of $M_{ij}$. 

The mean NDCG@10 for held out queries in the dataset is assumed to be ground truth. We define an $|R| \times |R|$ preference matrix $P$ in which $P_{ij}$ is $1$ if ranker $R_i$ has a higher NDCG@10 than ranker $R_j$, 0 if $R_j$ has a higher score than $R_i$, and $0.5$ if the two rankers have the same score.

For a given pair of rankers, we consider the multileaving method to have made an error after $t$ comparisons if $\hat{M}_{ij}(t)$ and $P_{ij}$ are not equal. We wish to minimize the percentage of errors made,
\begin{equation}
E(t)=\frac{\sum_{i,j \in R}sgn(\hat{M}_{ij}(t)-0.5) \neq sgn(P_{ij}-0.5)}{|R|(|R|-1)}
\label{eq:groundtruth}
\end{equation}

For these experiments we compare feature rankers from the MSLR-WEB30k \cite{letor2012}, YLR1 and YLR2 \cite{chapelle2011yahoo} datasets. Feature rankers can be rankers like PageRank or the BM25 score of the body of a document. Feature rankers were also used for the experimental setup in \cite{schuth2014multileaved,schuth2015probabilistic}. 

We use a simulated user setup. For each iteration we randomly sample with replacement a query from the pool of queries of the dataset. The rankers being compared are then multileaved, and clicks on the multileaved list are generated from three different probabilistic user models: the perfect, navigational and informational click models as described in \cite{hofmann2013fidelity}.

For each dataset the queries are split into a training and test set. For a given run on $k$ rankers, we randomly sample $k$ feature rankers from the given dataset and compute NDCG@10 for each of these rankers on the test query set. These NDCG@10 scores are used to create a ground truth preference matrix $P$ against which we can measure the percentage errors $E$ defined in Equation~\ref{eq:groundtruth}. For each iteration we then randomly sample a query from the training set, multileave the $k$ rankers using each multileaving method, and compute $E$ for each method. We then investigate how $E$ develops at each iteration. Additionally we show the percentage errors of the NDCG@10 score computed only from the queries so far used for multileaving. This serves as a lower bound on the error that can reasonably be obtained. For PM we fix the sample size parameter at 10,000 as in \cite{schuth2015probabilistic}.

\noindent\textbf{Findings:} 

Table~\ref{Table:results} enumerates the percentage error after 2,000 and 10,000 iterations when multileaving 5, 40 or 100 rankers for the three click models. In almost all cases SOSM is superior. The two exceptions occur when comparing only 5 rankers. In this case, TDM is marginally better than SOSM for the informational click model, and equivalent for the perfect click model. For 40 or 100 rankers, SOSM substantially outperforms TDM and PM. For example, with 100 rankers and the informational click model, the error rates are reduced by 50\% from 32\% to 16\% after 10,000 iterations.

\begin{table*}[ht]
\caption{Percentage error, $E$, after 2,000 and 10,0000 iterations for each multileaving method for 5, 40 and 100 rankers and three click models. The best performing method is bolded and * indicates a statistically significant difference with $p<0.01$ to both the baseline methods according to paired t-tests.}
\begin{center}
\begin{tabular}{ | c | c || c | c | l | c | c | l |}
  \hline
   & Iterations & \multicolumn{3}{|c|}{2,000} & \multicolumn{3}{|c|}{10,000}\\ \hline \hline
   Click Model &  \backslashbox{\# Rankers}{Method} & TDM & PM & SOSM & TDM & PM & SOSM \\ \hline \hline
   \multirow{3}{*}{Perfect} & 5 rankers & \textbf{18\%} & 23\% & \textbf{18\%} & 14\% & 18\% & \textbf{14\%} \\ \cline{2-8}
   & 40 rankers & 23\% & 28\% & \textbf{17\%*} & 18\% & 28\% & \textbf{15\%*} \\ \cline{2-8}
   & 100 rankers & 30\% & 29\% & \textbf{18\%*} & 21\% & 28\% & \textbf{16\%*} \\ \hline
   \multirow{3}{*}{Navigational}  & 5 rankers & 27\% & 30\% & \textbf{22\%} & 24\% & 25\% & \textbf{16\%} \\ \cline{2-8}
   & 40 rankers & 34\% & 30\% & \textbf{24\%*} & 26\% & 31\% & \textbf{22\%*} \\ \cline{2-8}
   & 100 rankers & 39\% & 31\% & \textbf{25\%*} & 29\% & 29\% & \textbf{21\%*} \\ \hline
   \multirow{3}{*}{Informational} & 5 rankers & \textbf{18\%} & 29\% & 20\% & \textbf{16\%} & 32\% & 18\% \\ \cline{2-8}
   & 40 rankers & 37\% & 32\% & \textbf{22\%*} & 27\% & 32\% & \textbf{15\%*} \\ \cline{2-8}
   & 100 rankers & 42\% & 33\% & \textbf{21\%*} & 34\% & 32\% & \textbf{16\%*} \\ \hline 
\end{tabular}
\label{Table:results}
\end{center}
\end{table*}

\begin{figure*}[ht]
  \centering
  \subfloat[][perfect click model]{
  \includegraphics[width=0.32\textwidth]{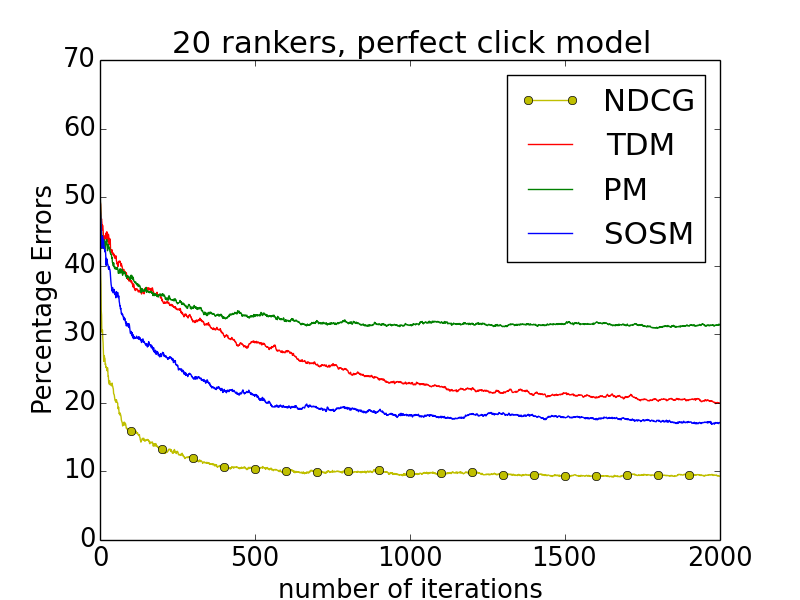}  
  }
  \subfloat[][navigational click model]{
    \includegraphics[width=0.32\textwidth]{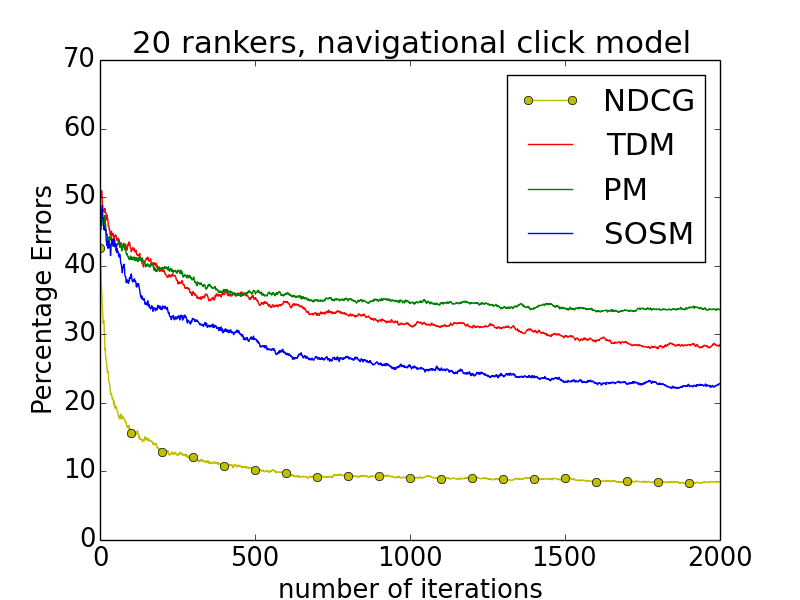}  
  }
  \subfloat[][informational click model]{
    \includegraphics[width=0.32\textwidth]{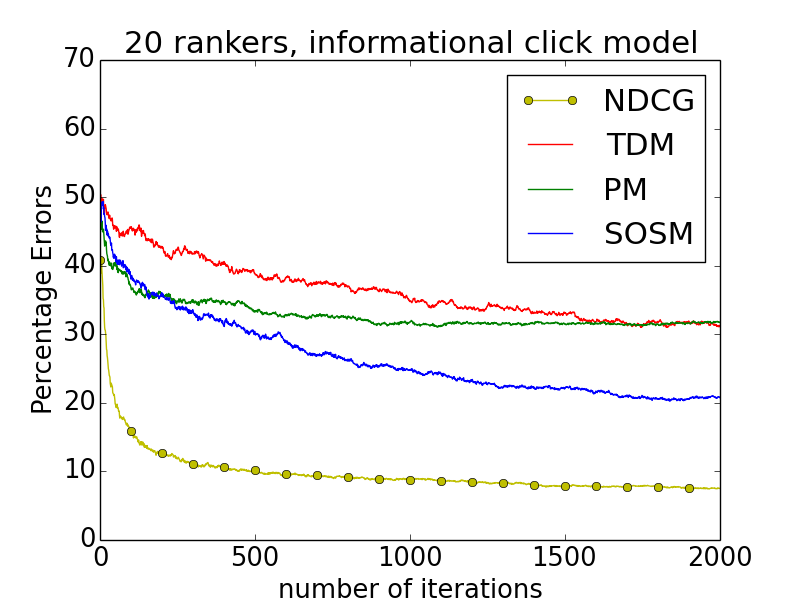}  
  } 
\caption{Percentage errors (averaged over 25 runs) versus the number of iterations on random subsets of $20$ rankers for the MSLR dataset using perfect (a), navigational (b), and informational (c) click models.}
\label{fig:ClickModels}
\end{figure*}

Figure~\ref{fig:ClickModels} shows how the performances of the multileaving methods are affected by the click model used. For all three click models, SOSM outperforms both TDM and PM. This is most pronounced for the navigational model, where the percentage error for TDM and PM is 50\% greater than that of SOSM (30\% compared to 20\%).

\begin{figure}[ht]
  \centering
  \subfloat[][YLR1]{
    \includegraphics[width=0.65\columnwidth]{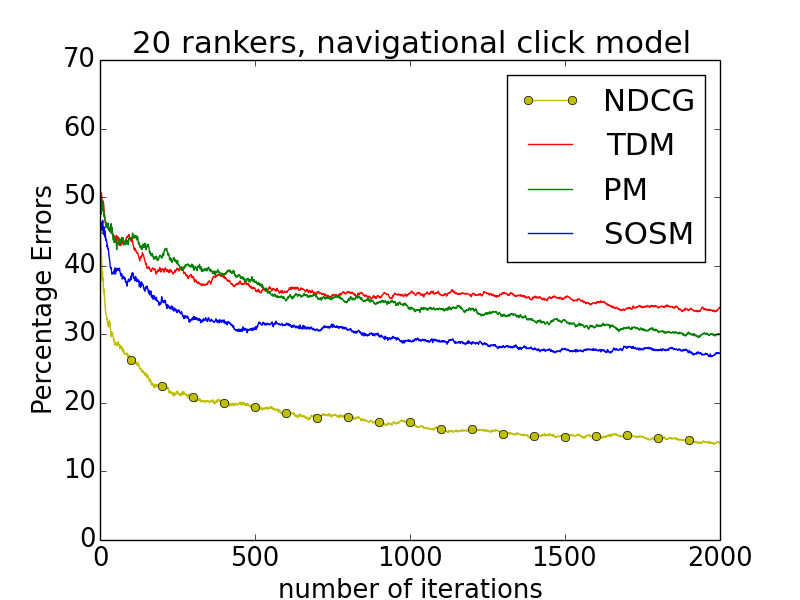}  
  } \\
  \subfloat[][YLR2]{
    \includegraphics[width=0.65\columnwidth]{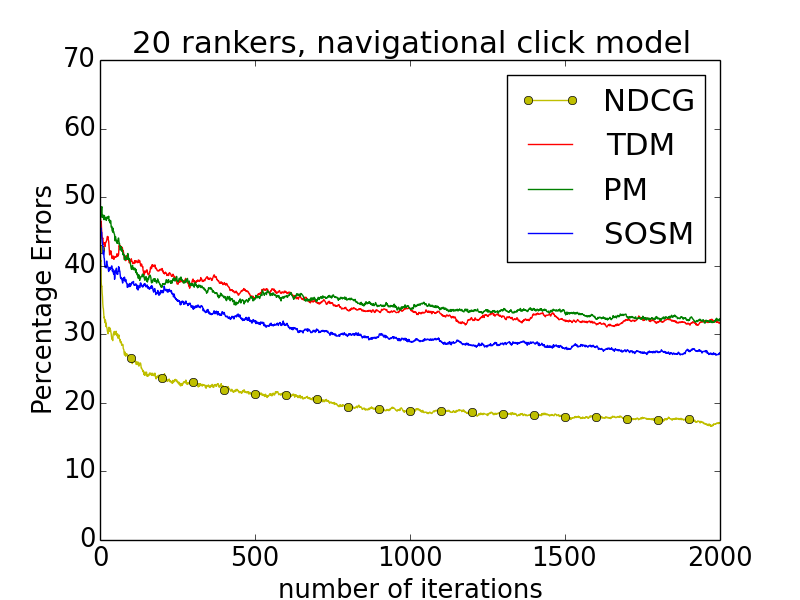}  
  } 
\caption{Percentage errors (averaged over 25 runs) versus the number of iterations on random subsets of $20$ rankers for the YLR1 (a) and YLR2 (b) datasets using a navigational click model.}
\label{fig:Datasets}
\end{figure}

Figure~\ref{fig:Datasets} shows the sensitivity of the multileaving methods to different choices of dataset. We repeat the experiment from Figure~\ref{fig:ClickModels}(b) using two other datasets.  All three algorithms (TDM, PM, SOSM) perform similarly across datasets. In all cases, SOSM exhibits superior performance.

\begin{figure*}[ht!]
  \centering
  \subfloat[][perfect click model]{
  \includegraphics[width=0.32\textwidth]{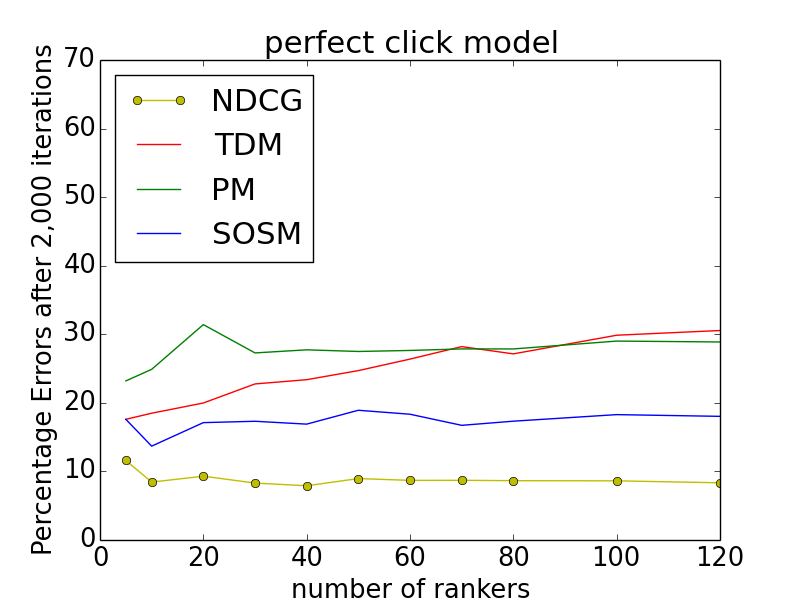}  
  }
  \subfloat[][navigational click model]{
    \includegraphics[width=0.32\textwidth]{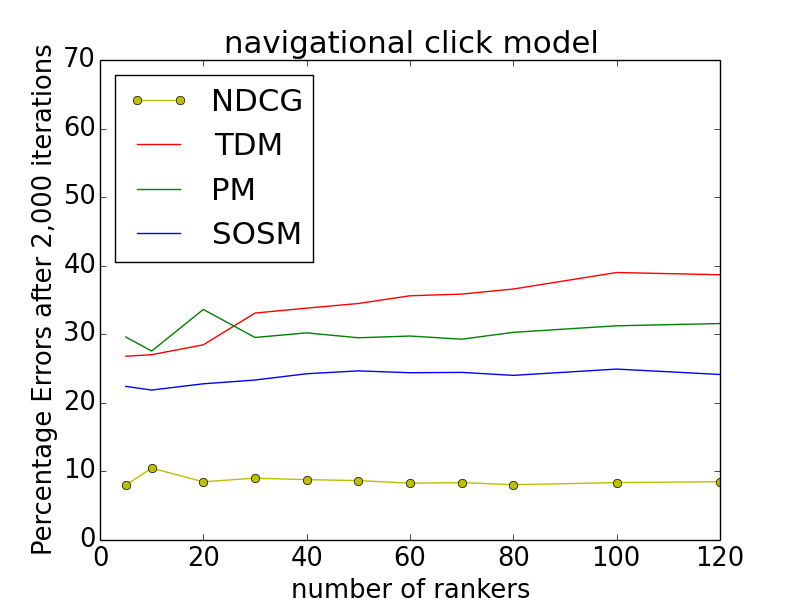}  
  }
  \subfloat[][informational click model]{
    \includegraphics[width=0.32\textwidth]{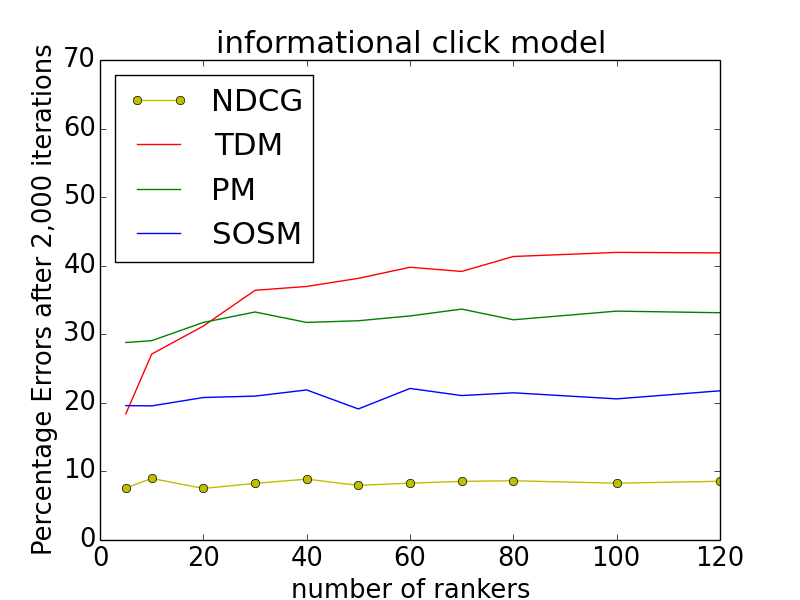}  
  } 
\caption{Percentage errors after 2,000 iterations (averaged over 25 runs) versus the number of rankers being compared on 25 random subsets of $k$ rankers for the MSLR dataset with perfect (a), navigational (b), and informational (c) click models.}
\label{fig:VaryRankers}
\end{figure*}

Figure~\ref{fig:VaryRankers} shows how the performances of the multileaving methods vary with the number of rankers being compared. We show the percentage error after 2,000 iterations, as a function of the number $k$ of rankers that are multileaved. For a given $k$, a random subset of rankers is selected and multileaved for 2,000 iterations. The same subset is used for PM, TDM and SOSM. This is repeated 25 times, each time with a different random subset of $k$ rankers.  The results in Figure~\ref{fig:VaryRankers} are the average of these 25 runs. We observe that for SOSM and PM the error remains relatively stable as the number of rankers increases. However, for TDM the error is increasing and we observe that its performance becomes worse than PM for large numbers of rankers. This is due to the fact that during the scoring phase, TDM is unable to assign credit to more than the 10 rankers from which the multileaved documents originated. 

\begin{figure}
  \centering
  \subfloat[][random click model]{
  \includegraphics[width=0.65\columnwidth]{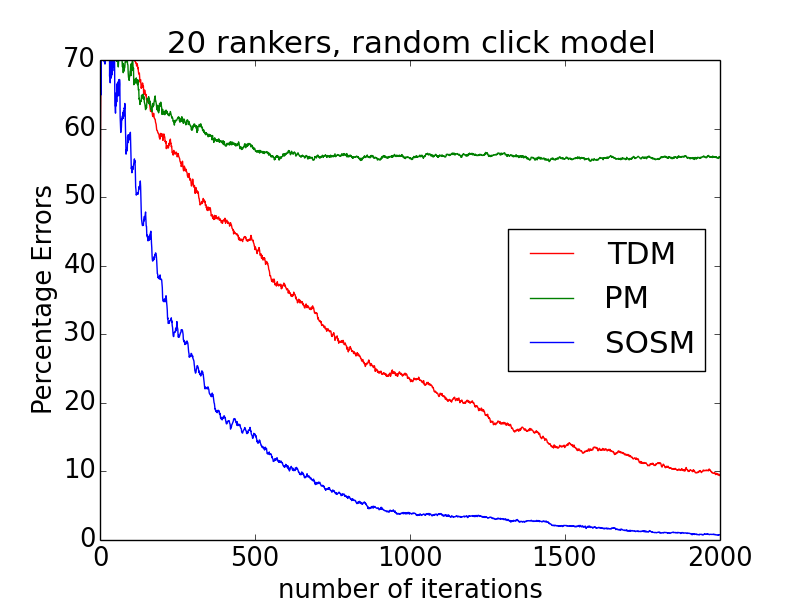}  
  } 
\caption{Percentage errors (averaged over 25 runs) versus the number of iterations on random subsets of $20$ rankers for the MSLR dataset using a random click model.}
\label{fig:bias}
\end{figure}

Figure~\ref{fig:bias} tests if the multileaving methods are biased. In Section~\ref{sec:related}, we showed that PM could exhibit bias under certain conditions. For this experiment we use a random click model, i.e. clicks are random and independent of document relevance. In this case, we expect that the elements of the pairwise preference matrix should converge to 0.5, i.e. there is no observed preference between rankers $i$ and $j$. In this case, an error is declared if the value of $\hat{M}_{ij}(t)$ deviates from 0.5 by more than 0.03. Figure~\ref{fig:bias} shows that PM exhibits very strong bias, with error rates of about 60\%.\footnote{Note that in \cite{schuth2015probabilistic} no such bias was detected. However, in \cite{schuth2015probabilistic} the set of multileaved rankers was not picked randomly. Further, personal communication with an author of \cite{schuth2015probabilistic} confirmed the existence of a software bug in PM which we corrected for these experiments.}
TDM's behaviour is much better, but after 2000 iterations some bias is still present. In contrast, the percentage error decreases much quicker in SOSM, and is almost zero after just 2000 iterations.

\section{Conclusion and Future Work}

We identified and experimentally verified weaknesses in the scalability of TDM and the unbiasedness of PM. We then proposed a new algorithm, SOSM, that corrects these problems. Experimental results using simulated users (perfect, navigational, informational click models), on three different datasets confirmed that (i) SOSM scales well with the number of rankers to be multileaved, (ii) is unbiased, and (iii) has significantly less error than prior methods. In some cases error rates were reduced by half.

The residual error needs investigating but is likely to be partly due to (i) establishing a ground truth based on NDCG@10, which is not used as the scoring function in Equation~\ref{eq2}, and (ii) the ground truth data is computed on ``test'' data that is not used during the multileave experiments.

Future work will investigate scoring functions allowing the multileaving method to optimise for specific evaluation measures such as NDCG@10 or to agree with specific measures of user satisfaction.

\bibliographystyle{abbrv}


\begin{thebibliography}{1}
\scriptsize
\bibitem{hofmann2013fidelity}
K.~Hofmann, S.~Whiteson, and M.~D. Rijke.
\newblock Fidelity, soundness, and efficiency of interleaved comparison
  methods.
\newblock {\em TOIS}, 31(4):17.1-17.39, 2013.

\bibitem{schuth2015probabilistic}
A.~Schuth, R.-J. Bruintjes, F.~B{\"u}ttner, J.~van Doorn, C.~Groenland,
  H.~Oosterhuis, C.-N. Tran, B.~Veeling, J.~van~der Velde, R.~Wechsler, et~al.
\newblock Probabilistic multileave for online retrieval evaluation.
\newblock \emph{SIGIR}, pages 955--958, 2015.

\bibitem{schuth2014multileaved}
A.~Schuth, F.~Sietsma, S.~Whiteson, D.~Lefortier, and M.~de~Rijke.
\newblock Multileaved comparisons for fast online evaluation.
\newblock \emph{CIKM}, pages 71--80. 2014.

\bibitem{chapelle2012large}
O.~Chapelle, T.~Joachims, F.~Radlinski and Y.~Yue, and M.~de~Rijke.
\newblock Large-scale validation and analysis of interleaved search evaluation.
\newblock \emph{TOIS}, 30(1):6.1-6.41, 2012.

\bibitem{chapelle2011yahoo}
O.~Chapelle and Y.~Chang.
\newblock Yahoo! Learning to Rank Challenge Overview.
\newblock \emph{JMLR}, 14:1-24, 2011

\bibitem{letor2012}
\newblock Microsoft Learning to Rank Datasets, 2012
\newblock http://research.microsoft.com/en-us/projects/mslr/default.aspx

\end{thebibliography}

\end{document}